\documentclass[12pt]{article}
\usepackage{graphics,color}
\newcommand{\be}{\begin{equation}}
\newcommand{\ee}{\end{equation}}
\newcommand{\bea}{\begin{eqnarray}}
\newcommand{\eea}{\end{eqnarray}}
\def\ba{\begin{array}} 
\def\ea{\end{array}}

\def\rb{{\bf r}}

\title{Bose condensation in two dimensions with disorder:
Gross-Pitaevskii approach}
\author{
Dmitri~E.~Nikonov
and Ata\c{c}~Imamo\u{g}lu \\ \em
Department of Electrical and Computer Engineering \\
\em and Center 
for Quantized Electronic Structures (QUEST), \\
\em University of California, Santa Barbara, California 93106-9560}

\begin{document} 

\maketitle

\abstract
Bose condensation of interacting bosons 
in a two-dimensional random potential is 
studied. The Gross-Pitaevskii equation is solved to 
determine the spatially-varying order parameter
and the localization length as a function of the 
disorder, the interaction strength, and 
the condensate density. 
A finite temperature of condensation is obtained 
thereby. The results are applied to determination of 
the superradiant decay of excitons in a GaAs quantum well.
\endabstract

\centerline{PACS number(s): 03.75.Fi, 71.23.An, 75.40.Cx}


Bose condensation in a disordered medium was the topic of extensive
research starting a decade ago \cite{ma86,fisher88}.
It was mainly concerned with the superfluidity of 
liquid $^4$He in porous media or on substrates
\cite{fisher89,lee90,krauth91}
or with the superconductor-insulator transition
in high critical temperature superconductors
\cite{ma85,fisher90,gold92,nelson92}.
In most cases, the Hubbard Hamiltonian of 
interacting bosons on a lattice is employed.
This model is treated 
via a mean-field theory \cite{sheshadri93},
via the Bogolyubov approximation \cite{singh94},
in the approximation of Gaussian fluctuations \cite{nisam93},
by the Monte Carlo numerical methods
\cite{scalettar91,makivic93,onogi94},
or by renormalization group methods
\cite{singh92,pai96}.

Bose condensation in other systems is often treated via the 
Gross-Pitaevskii equation (nonlinear Schr\"odinger equation)
for the condensate \cite{lifshitz80}.
The advantages of this approach are that
one immediately obtains 
the order parameter of a non-uniform condensate and 
that the non-physical 
features related to the integer occupancy number
of a lattice site do not appear. 
The connection between two approaches was made in 
Refs.~\cite{lee90,singh94}.

Our interest in 
Bose condensation in a
 two-dimensional space is primarily motivated by 
the ongoing experiments aimed at observing quantum degeneracy
of excitons in quantum wells
\cite{butov94prl}. Since in two or smaller dimensions
the critical temperature of condensation would be 
zero in a uniform infinite space, only localization by
geometrical size, or, in our case, localization by disorder
can cause condensation at a finite temperature.
This phase, consisting of a number of separate and presumably 
phase-uncorrelated ``lakes'' of Bose condensate is more properly 
termed ``Bose glass'' \cite{fisher89}.

For excitons, the order parameter is proportional
to the polarization of the medium. Therefore Bose condensation
has a profound influence on the optical properties of 
excitons, especially on the superradiant decay.
The smallest of the exciton mean free path
and the localization length gives the coherence area of 
an exciton. It was shown in 
Ref.~\cite{rashba62,feldmann87,citrin93}
that the coherence area is related to the linewidth and 
the radiative lifetime  of excitons.
Bjork et al. discussed it in terms of superradiance of excitons 
\cite{bjork94}.
A sharp increase of the exciton oscillator strength was suggested 
as a criterion of Bose condensation of excitons 
\cite{butov94prl,butov96}. The oscillator strength reaches 
saturation as the coherence length approaches the wavelength.
Other approaches (Gerrit Bauer, private communication)
predict a divergence of the coherence area as the temperature 
decreases down to some critical temperature, and a constant
coherence area below this temperature.
In view of this controversy, it is important to give 
an intuitive treatment of the oscillator strength and 
the superradiant decay.

In this paper we numerically solve the Gross-Pitaevskii equation  
for the order parameter of the condensate
in a disordered two-dimensional medium modeled by a set of potential
wells with their depths depending on the widths.
[The authors understand the limited nature of such an assumption,
but stress that such a potential might be applicable to 
quantum dots defined by the variation of a quantum-well width].
In this way we determine the localization size 
as a function of the disorder strength, the interaction strength, and 
the condensate density. This relation along with the 
integral of the Bose distribution 
for a localized condensate allow us to obtain
the temperature of condensation for a given density of bosons.
The localization length determines the oscillator strength 
and the increase of the rate of superradiant decay.


We describe the condensate of interacting bosons via the 
stationary
Gross-Pitaevskii equation for the long-range order parameter
$\psi$ \cite{lifshitz80}
\be
-\frac{\hbar^2}{2M} \nabla^2 \psi + V(\rb)\psi + U\psi|\psi|^2 = 
\mu\psi,
\label{GPini}
\ee
where $M$ is the mass of a boson, $U$ is the interaction strength,
$\mu$ is the chemical potential, and $V$ is the external potential.
This model implies that we take the lifetime of the boson particles to
be infinite and neglect scattering.

This would be a good approximation for indirect excitons 
(having the electron and the hole in different quantum wells)
in a direct-bandgap semiconductors. Equation (\ref{GPini})
also implies that the exciton-exciton interaction can be modeled 
by a short-range $s$-wave scattering. Even though this assumption 
is at best questionable, we make it for the sake of simplicity.

To incorporate  a random potential, caused, 
e.g., by impurities or fluctuations
of the quantum well thickness, we adopt the following model.
The whole two-dimensional 
space is divided into a set of ``lakes'' of condensate of 
the characteristic size $L_c$.
The condensate in each of them is presumed to be trapped by
a cylindrical potential well of a radius $L$ to be determined:
following Ref.~\cite{ziman79} we choose the depth of this well
to be inversely proportional to its radius
\be
V(\rb) = - \frac{\xi}{L} \theta(L-|\rb|),
\ee
where $\xi$ is the parameter characterizing the strength of 
the disorder.
The average density of bosons in the condensate 
is $n_c$. 
The average number  of bosons in each lake is 
$N_c = n_c L_c^2$. 
We impose a condition to normalize the 
number of bosons in each condensate lake
\be
\int |\psi|^2 d\rb = n_cL_c^2, 
\label{int1} 
\ee
and 
we identify 
$L_c$ with the coherence length defined by 
the following relation
\be
\int |\psi|^4 d\rb = n_c^2L_c^2. 
\label{int2}
\ee
The integrals have infinite limits.

In order to reveal the self-similarity of the solutions,
let us introduce a constant length scale $L_0$ and correspondingly
express all densities in terms of $1/L_0^2$. 
On the other hand, the coordinates 
will be expressed in terms of  the well width 
$L$ and the wavefunction will be expressed in terms of  
$1/L$. Then the Gross-Pitaevskii equation in dimensionless variables
becomes
\be
-\nabla^2 \psi -\xi_0 L \theta(1-|\rb|)\psi + u\psi|\psi|^2 = 
\mu_0 L^2\psi,
\label{GPdim}
\ee
where the dimensionless parameters
\bea
\xi_0 = \frac{2 M \xi L_0}{\hbar^2}, \\
u = \frac{2 M U}{\hbar^2}, \\
\mu_0 = \frac{2 M \mu L_0^2}{\hbar^2}.
\eea
The parameter $l_h$ is called ``healing length'' of the condensate,
\be
l_h^2 = \frac{2\hbar^2}{MUn_c} = \frac{4}{un_c}.
\ee
By comparing the first and the third terms in (\ref{GPdim}),
one can estimate that the order parameter can appreciably change
only over the dimensionless length of the order of $l_h$.
If $u$ is large, the healing length is 
less than the interparticle separation, and if 
$u$ is small, it is larger that the interparticle separation.
Under this change of variables (\ref{int1}) and
(\ref{int2}) become
\bea
\int |\psi|^2 d\rb = n_cL_c^2, 
\label{int1dim} \\
\int |\psi|^4 d\rb = n_c^2L_c^2L^2. 
\label{int2dim}
\eea
At any given strength of disorder $\xi$, for convenience,
we chose the length scale 
\be
L_0 = \frac{\hbar^2}{2M\xi}
\ee
so that $\xi_0 = 1$.
Besides, we notice a self-similarity of the set of the equations 
(\ref{GPdim})-(\ref{int2dim}) under the transformation with an
arbitrary positive $a$
\be
u \rightarrow u' = u/a, \quad n \rightarrow n' = a n,
\quad \psi \rightarrow \psi' = \sqrt{a} \psi.
\label{self}
\ee
Therefore we numerically solve the set of equations 
for $u=1$ and obtain the general solution by rescaling
according to (\ref{self}).

\begin{figure} 
\begin{center}
\rotatebox{0}{\scalebox{1}{\includegraphics{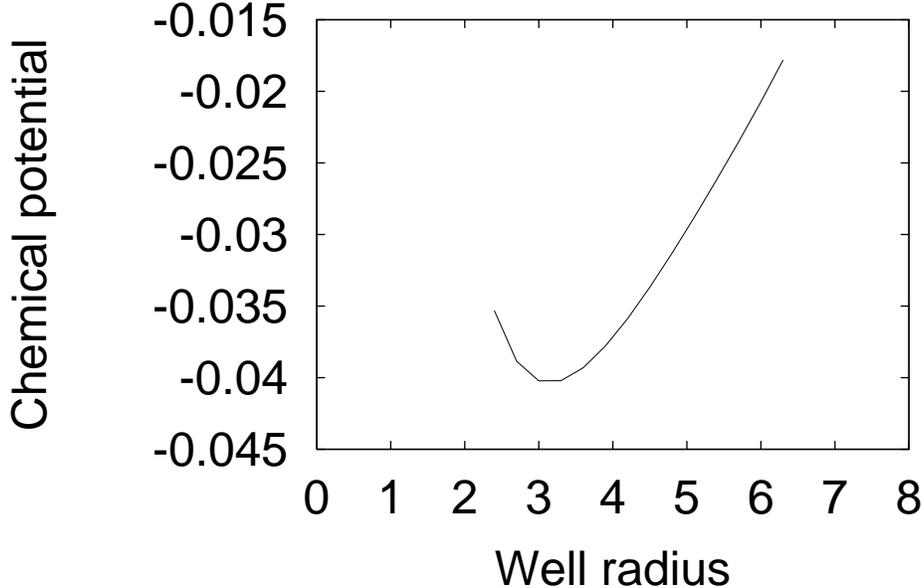}}}
\end{center}
\caption{Dependence of the chemical momentum of the condensate on 
the trapping well width at $u=1$, $n=0.06$.}
\label{fi:minim}
\end{figure}

The solution of the dimensionless equations 
(\ref{GPdim}) - (\ref{int2dim}) determines
the dependence of the chemical
potential of the condensate on the trapping radius $L$.
An example of it for $u=1$ is shown in Fig.~\ref{fi:minim}.
For the trapping radii outside the region on the plot 
no localized solutions (with $\mu<0$) 
 were found. 
At $u=0$, a linear Schr\"odinger equation, 
 solutions exist for any radius $L$ of the well. As the density of 
bosons, or, equivalently, the nonlinear coefficient $u$,
increases, the interaction of
bosons screens the random potential more strongly \cite{lee90}.
As a result,
the range of $L$, where localized solutions are 
possible, narrows until 
it totally disappears.
The radius $L$ is determined from the solution of (\ref{GPini})
which minimizes the chemical potential $\mu$.
Note that extended condensate solutions are
not a limit of localized solutions.
The former have a positive chemical potential $\mu = un_c$.
The latter must have negative $\mu$ to ensure decrease 
$\psi \to 0$ at $|x| \to \infty$.

\begin{figure} 
\begin{center}
\rotatebox{0}{\scalebox{1}{\includegraphics{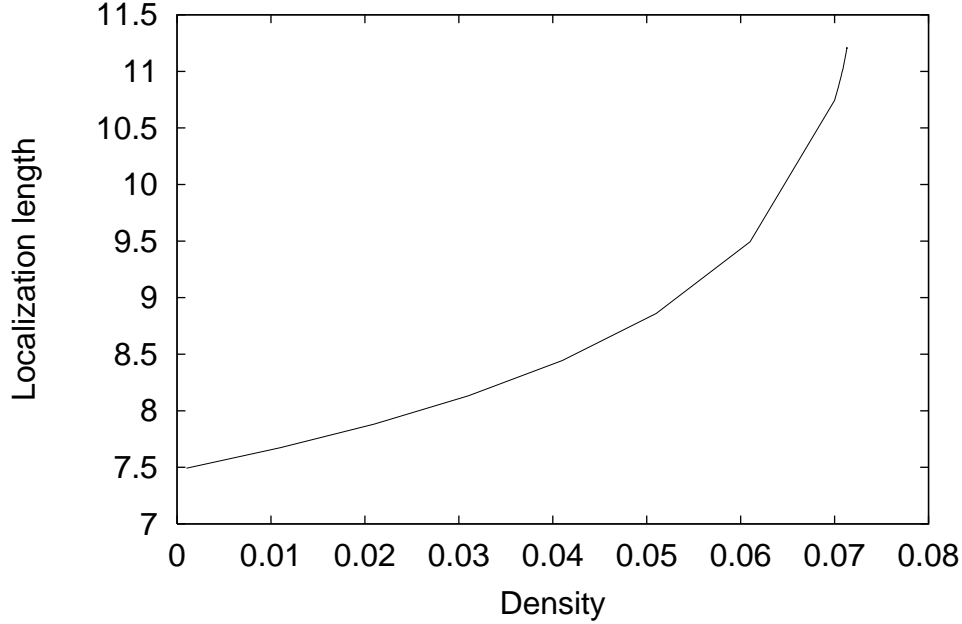}}}
\end{center}
\caption{Dependence of the localization length of the condensate 
on its density at $u=1$.}
\label{fi:crit}
\end{figure}

Based on these solutions, we seek the dependence of 
the condensate size $L_c$ on 
the average condensate density $n_c$.
The numerical result  
for $u=1$
is shown in Fig.~\ref{fi:crit}.
With a good accuracy at all densities this dependence
can be approximated by
\be
L_c = \alpha (n_g - n)^\beta
\label{crit}
\ee
with $\alpha = 5.4$, $\beta = -0.1317$, and $n_g = 0.074$.
Thus below the critical density (proportional to 
the dimensionless constant 
$n_g$), the condensate exists as a Bose glass
with localized ``lakes'' of condensate, and, above this density,
an extended Bose condensate is possible at zero temperature.
This universal dependence will be used later while treating
bosons with different nonlinear coefficient $u$ and at non-zero
temperatures.


Now we apply our model to the case of excitons in GaAs
quantum wells.
Let us estimate the interaction strength $U$ for excitons
in the ideal two-dimensional case.
It is proportional to the exciton-exciton interaction term, i.e.,
the Coulomb exchange integral 
in the exciton 
ground (1s) state, which, according to \cite{schmitt89}
is equal to
\be
I = \sum_{kk'}[\phi_{1s}(k')-\phi_{1s}(k)]
\phi^*_{1s}(k) |\phi_{1s}(k)|^2 V(k-k') = 
4\pi a_0^2 (1 - \frac{315 \pi^2}{4096}) E_0,
\label{Iexcit}
\ee
where $\phi_{1s}(k)$ is the wavefunction of the 1s exciton
state and $V(k)$ is the Coulomb potential 
in the momentum representation,
$a_0$ is the Bohr radius and $E_0$ is the binding energy of an exciton
in two dimensions given in \cite{schmitt89}. 
We have neglected screening of the Coulomb potential by other
excitons, which is possible to do for small densities 
(interparticle separation much larger than $a_0$).
At higher densities (\ref{Iexcit}) overestimates the strength
of interaction $u$.

The estimate of $I$ for indirect excitons will be inserted here.

Therefore we obtain that
\be
u = \frac{4 M I}{\hbar^2} \approx \frac{6.06 M}{m_r},
\ee
bearing in mind that $M = m_e+m_h$ and $m_r=m_em_h/M$.
For the case of GaAs, where $m_e = 0.0665m_0$, $m_h = 0.377m_0$
and $m_0$ is the mass of a free electron, we obtain $u \approx 47$,
which means strong boson-boson interaction.
It is the opposite of Bose condensation in atomic traps
in which one encounters a weak interaction,
\be
u \sim \frac{a_S}{L_{\rm trap}},
\ee
where $a_S$ is the scattering length in $s$-wave collisions,
and $L_{\rm trap}$ is the size of the trap.


It is well-known that the critical temperature of 
Bose condensation in two dimensions  is zero for an infinite
space. Therefore
only a localized (Bose glass) rather than an extended condensate 
can exist at non-zero temperatures.
Only a fraction (if any) of bosons
will be in the condensate and be described by the Gross-Pitaevskii 
equation. 
\footnote{We disregard the possibility of the 
Kosterliz-Thouless transition, see e.g. \cite{stoof93}.}
We need to add the relations describing the statistics
of bosons outside the condensate. Following \cite{ketterle96}
we obtain for a two-dimensional localized Bose gas with the 
average density $n$
\bea
1 - \frac{n_c}{n} = \frac{\Lambda_{cr}^2}{\Lambda^2},
\label{sys1} \\
n \Lambda_{cr}^2 = \log\left( \frac{2 L_c^2}{\Lambda_{cr}^2}  \right),
\label{sys2}
\eea
where $k_B$ is the Boltzmann constant,
the de Broglie wavelength corresponding to the temperature $T$ is 
\be
\Lambda = \sqrt{ \frac{2\pi\hbar^2}{Mk_B T}   },
\ee
and $\Lambda_{cr}$ is the wavelength corresponding to a critical
temperature for the given localization size $L_c$.
Here we again assume the same scaling of lengths and densities.
Equations (\ref{sys1}) and (\ref{sys2}) together with the rescaled
(\ref{crit})
\be
L_c = \alpha (n_g - u n_c)^\beta
\label{sys3}
\ee
allow us to determine the condensate density $n_c$ and the localization
length $L_c$ for the given concentration and temperature.
We see that the condensate density cannot exceed the critical value
$n_g/u$.

The correlation length allows us to obtain the collective 
dipole of the condensate and the rate of radiative decay.
Exciton condensate is similar to a collective of excitons 
excited by a laser. The differences are: 1) that the direction and 
the phase of the dipole are not given by the laser, but 
undetermined until a measurement is made on the condensate;
2) the order parameter does not contain a spatial dependence
associated with the wavevector of the exciting laser.
With these modifications, we apply the results of Bjork et al.
\cite{bjork94}. 
The number of modes available for excitons 
are 
\be
N_e = \frac{8 a_c^2}{a_0^2},
\ee
where the condensate radius is such that $\pi a_c^2 = L_c^2$.
In the case of a thin quantum well and large $N_e$,
according to \cite{rehler71},
the power of emission per unit angle is proportional to
$I(\phi,\chi)\Gamma(\phi)$, where 
the usual dipole pattern
\be
I((\phi,\chi) = \cos^2\chi + \sin^2\chi\cos^2\phi,
\ee
and the factor of the collective emission is  
\be
\Gamma(\phi) = \left( 
\frac{ 2 J_1(ka_c \sin\phi)}{ka_c \sin\phi} \right)^2
\ee
Here $k = 2\pi/\lambda$, and $\lambda$ is the wavelength
in the material (228nm in GaAs),
$\phi$ is the angle between the direction of emission 
and the orthogonal to the well.
The cooperativity parameter, which determines how well
the radiators are localized, is
\be
\mu_c = \frac{3}{8 k^2 a^2_c}
\int_0^\pi \sin\phi d\phi (1+\cos^2\phi) 
\Gamma(\phi).
\ee
For small condensates ($k a_c \ll 1$) it tends to 1,
and for large condensates it tends to $3/(ka_c)^2$.
The rate of superradiant emission of excitons $\gamma$
is increased compared to the bulk rate of free-electron-hole 
recombination $\gamma_0$ by the enhancement factor 
\be
\gamma = \mu_cN_e \gamma_0.
\ee
For small condensates the emission rate grows as $a_c^2$
to the same extent in all directions.
As the size $a_c$ of the condensate grows, the enhancement
factor tends to a constant.
However the angular distribution becomes 
strongly peaked at $\phi = 0$ where it still grows as $a_c^2$.

\begin{figure} 
\begin{center}
\rotatebox{0}{\scalebox{1}{\includegraphics{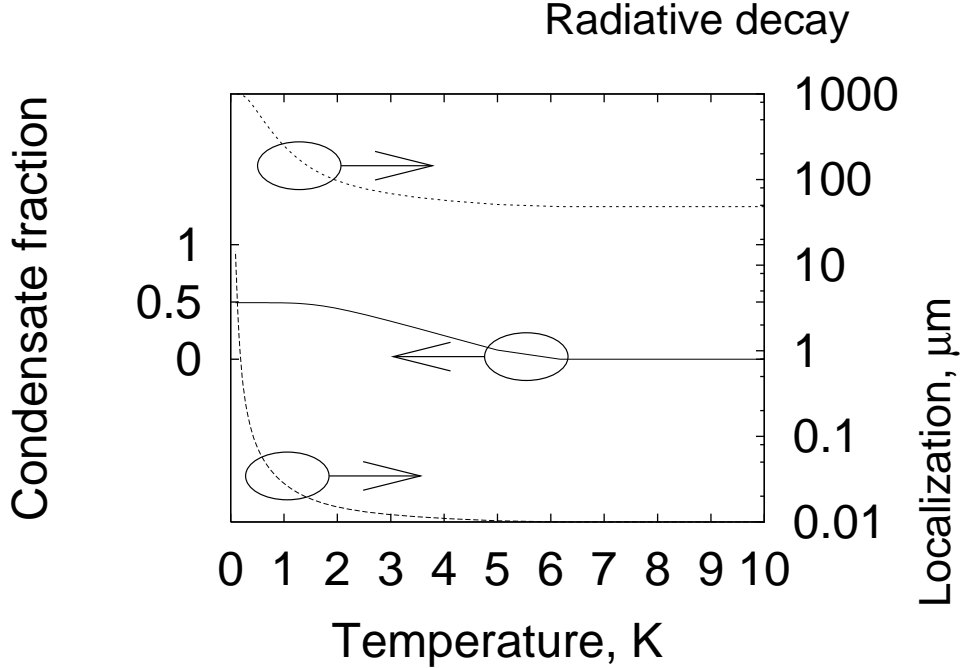}}}
\end{center}
\caption{Fraction of the bosons in the condensate
(middle curve)
the localization length in $\mu$m 
(lower curve)
and the factor of enhancement of radiative decay rate
(upper curve)
as functions of temperature
for $u=47$ and $L_0=10^{-8}m$: 
a) $n = 1.2 \times 10^{10}cm^{-2}$.}
\label{fi:tempera}
\end{figure}

\begin{figure} 
\begin{center}
\rotatebox{0}{\scalebox{1}{\includegraphics{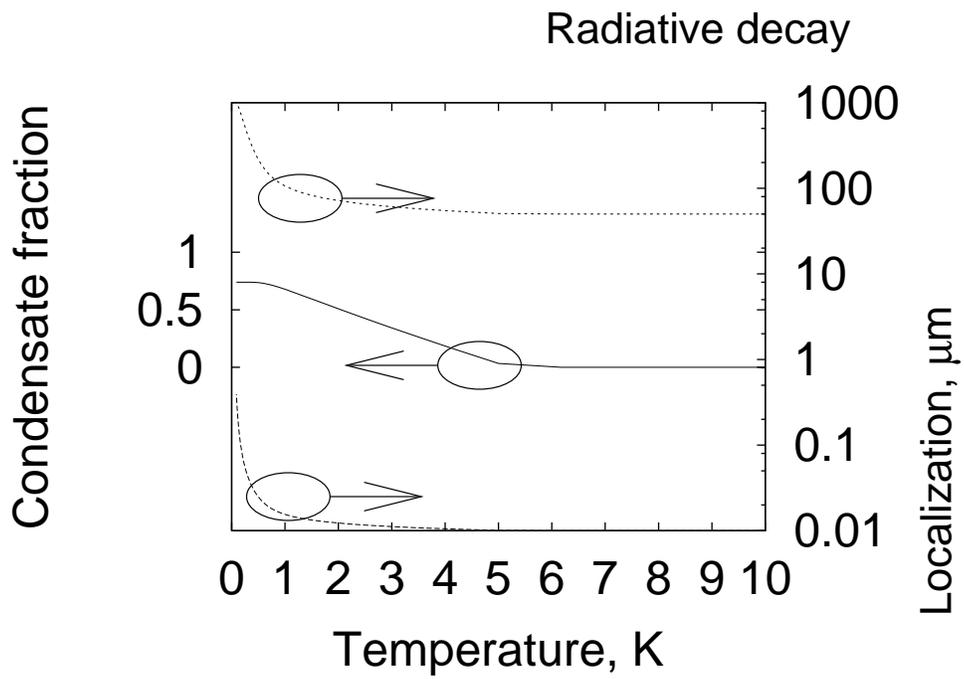}}}
\end{center}
\caption{Same as above: b) $n = 0.8 \times 10^{10}cm^{-2}$. }
\label{fi:temperb}
\end{figure}

\begin{figure} 
\begin{center}
\rotatebox{0}{\scalebox{1}{\includegraphics{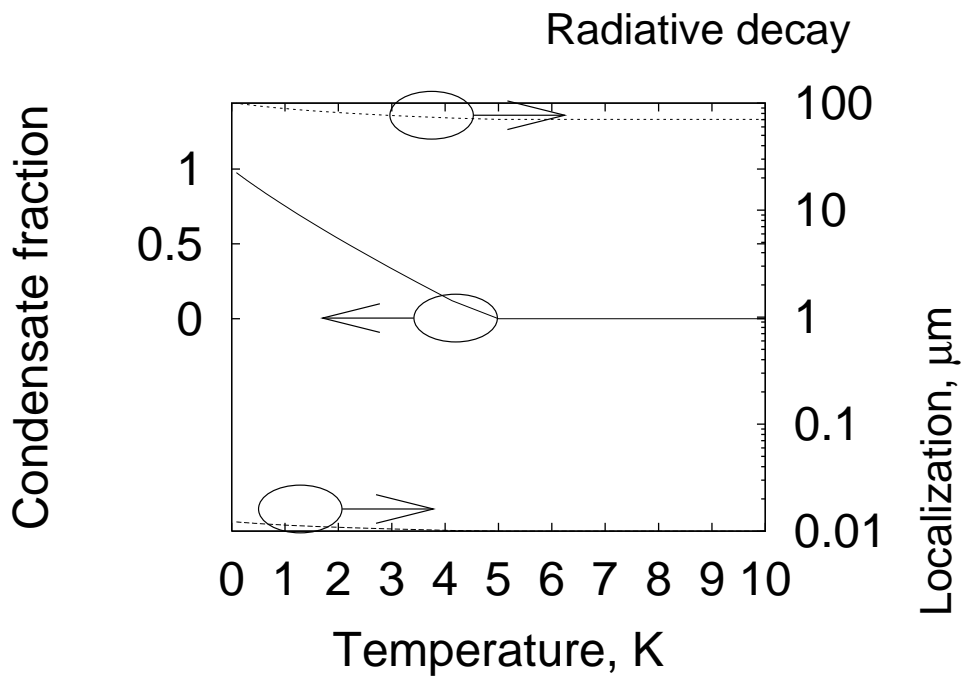}}}
\end{center}
\caption{Same as above: c) $n = 0.4 \times 10^{10}cm^{-2}$.}
\label{fi:temperc}
\end{figure}

The results of the numerical solution of (\ref{sys1})-(\ref{sys3})
are shown in Fig.~\ref{fi:tempera}-\ref{fi:temperc}.
For illustrative purposes we restore the dimension of length
by taking a specific disorder strength which corresponds to 
$L_0 = 10^{-8}$m.
We see that the condensate 
(a non-vanishing share of the condensate density)
appears below a well-defined temperature.
If the  total density of bosons is 
below the critical density, the fraction of 
the condensate tends to 1 as the temperature decreases, and 
the localization length tends to a small finite value,
see Fig.~\ref{fi:temperc}.
If the  total density is  above 
the critical one, the density of the condensate
tends to that critical value so that the condensate fraction
is always less than 1.
The localization length 
tends to infinity as the temperature decreases
and its value strongly depends on the total density,
see Fig.~\ref{fi:tempera}-\ref{fi:tempera}.
However this increase does not become noticeable until temperatures
several times below the condensation temperature.
The enhancement factor grows appreciably until it reaches the limiting
value as per reasoning above.


In conclusion, we have presented a theoretical description of 
Bose condensation in a random potential in two dimensions.
The disorder is believed to be sufficiently strong so that 
the Kosterliz-Thouless transition
the effects of the sample size can be disregarded.
We argued then that only localized condensates (Bose glass) can exist
at non-zero temperatures. 
A macroscopic occupation of the condensate
appears in a sharp transition below 
some condensation temperature. 
The average density in the condensate always stays 
below a certain critical density determined by the disorder strength.
As the temperature decreases, the localization length,
and consequently the coherence area, grow indefinitely if the 
average density of bosons is higher than the critical density.
The increase of the radiative decay rate reaches its limit
when the correlation length reaches the photon wavelength
in the medium. The increase in the rate 
of superradiant decay and, especially, in the power 
of emission orthogonal to the well, is a reliable indicator 
of Bose condensation.


This work is supported by  NSF, Packard Foundation, and
the Center for Quantized Electronic 
Structures (QUEST), an NSF Science and Technology Center.
We thank Leonid Butov for enlightening discussions and 
making available his experimental results before 
publication.




\begin{thebibliography}{10}

\bibitem{ma86}
M. Ma, B.~I. Halperin, and P.~A. Lee, Phys.~Rev.~B {\bf 34},  3136  (1986).

\bibitem{fisher88}
D.~S. Fisher and M.~P.~A. Fisher, Phys.~Rev.~Lett. {\bf 61},  1847  (1988).

\bibitem{fisher89}
M.~P.~A. Fisher, P.~B. Weichman, G. Grinstein, and D.~S. Fisher, Phys.~Rev.~B
  {\bf 40},  546  (1989).

\bibitem{lee90}
D.~K.~K. Lee and J.~M.~F. Gunn, J. Phys. {\bf 2},  7753  (1990).

\bibitem{krauth91}
W. Krauth, N. Trivedi, and D. Ceperley, Phys.~Rev.~Lett. {\bf 67},  2307
  (1991).

\bibitem{ma85}
M. Ma and P.~A. Lee, Phys.~Rev.~B {\bf 32},  5658  (1985).

\bibitem{fisher90}
M.~P.~A. Fisher, G. Grinstein, and S.~M. Girvin, Phys.~Rev.~Lett. {\bf 64},
  587  (1990).

\bibitem{gold92}
A. Gold, Physica C {\bf 190},  483  (1992).

\bibitem{nelson92}
D.~R. Nelson and V.~M. Vinokur, Phys.~Rev.~Lett. {\bf 68},  2398  (1992).

\bibitem{sheshadri93}
K. Sheshadri, H.~R. Krishnamurthy, R. Pandit, and T.~V. Ramakrishnan, Europhys.
  Lett. {\bf 22},  257  (1993).

\bibitem{singh94}
K.~G. Singh and D.~S. Rokhsar, Phys.~Rev.~B {\bf 49},  9013  (1994).

\bibitem{nisam93}
L.~Z. P.~Nisamaneephong and M. Ma, Phys.~Rev.~Lett. {\bf 71},  3830  (1993).

\bibitem{scalettar91}
R.~T. Scalettar, G.~G. Batrouni, and G.~T. Zimanyi, Phys.~Rev.~Lett. {\bf 66},
  3144  (1991).

\bibitem{makivic93}
M. Makivic\'{c}, N. Trivedi, and S. Ullah, Phys.~Rev.~Lett. {\bf 71},  2307
  (1993).

\bibitem{onogi94}
T. Onogi and Y. Murayama, Phys.~Rev.~B {\bf 49},  9009  (1994).

\bibitem{singh92}
K.~G. Singh and D.~S. Rokhsar, Phys.~Rev.~B {\bf 46},  3002  (1992).

\bibitem{pai96}
R. Pai, R. Pandit, H.~R. Krishnamurthy, and S. Ramasesha, Phys.~Rev.~Lett. {\bf
  76},  2937  (1996).

\bibitem{lifshitz80}
E.~M. Lifshitz and L.~P. Pitaevskii, {\em Statistical Physics, part 2}, Vol.~9
  of {\em Landau course}, 3rd ed. (Pergamon Press, Oxford, 1980).

\bibitem{butov94prl}
L.~V. Butov {\it et~al.}, Phys.~Rev.~Lett. {\bf 73},  304  (1994).

\bibitem{rashba62}
E.~I. Rashba and G.~E. Gurgenishvili, Sov. Phys. Solid State {\bf 4},  759
  (1962).

\bibitem{feldmann87}
J. Feldmann {\it et~al.}, Phys.~Rev.~Lett. {\bf 59},  2337  (1987).

\bibitem{citrin93}
D.~S. Citrin, Phys.~Rev.~B {\bf 47},  3832  (1993).

\bibitem{bjork94}
G. Bj{\"o}rk, S. Pau, J. Jacobson, and Y. Yamamoto, Phys.~Rev.~B {\bf 50},
  17336  (1994).

\bibitem{butov96}
B. L.~V,  in {\em The Physics of Semiconductors}, edited by R. Zimmermann
  (World Scientific, Singapore, 1996), p.\ 1927.

\bibitem{ziman79}
J.~M. Ziman, {\em Models of Disorder} (Cambridge University Press, Cambridge,
  1979).

\bibitem{schmitt89}
S. Schmitt-Rink, D.~S. Chemla, and D.~A.~B. Miller, Advances in Physics {\bf
  38},  89  (1989).

\bibitem{stoof93}
H.~T. Stoof and M. Bijlsma, Phys. Rev. E {\bf 47},  939  (1993).

\bibitem{ketterle96}
W. Ketterle and N.~J. van Druten, Phys.~Rev.~A {\bf 54},  656  (1996).

\bibitem{rehler71}
N.~E. Rehler and J.~H. Eberly, Phys.~Rev.~A {\bf 3},  1735  (1971).

\end{thebibliography}
\end{document}